\documentclass[
  reprint,
  superscriptaddress,
  amsmath,
  amssymb,
  aps,
  prl,
  longbibliography
]{revtex4-2}

\usepackage[utf8]{inputenc}
\usepackage[lining,semibold]{libertine}
\usepackage{amsthm}

\usepackage[libertine,cmintegrals,bigdelims,vvarbb]{newtxmath}

\usepackage{amsmath}
\usepackage{amsfonts}
\usepackage{bm}
\usepackage{mathrsfs}

\usepackage{graphicx}
\usepackage{array}[=2016-10-06]
\usepackage{dcolumn}
\usepackage{booktabs}
\usepackage{siunitx}

\usepackage{xcolor}
\definecolor{webgreen}{rgb}{0,.5,0}
\definecolor{webbrown}{rgb}{.6,0,0}
\definecolor{RoyalBlue}{rgb}{0.0,0.14,0.4}

\usepackage{hyperref}
\hypersetup{
  colorlinks=true,
  linktocpage=true,
  breaklinks=true,
  pdfstartpage=1,
  pdfstartview=Fit,
  pdfpagemode=UseOutlines,
  plainpages=false,
  bookmarksnumbered=true,
  bookmarksopen=true,
  bookmarksopenlevel=1,
  hypertexnames=true,
  pdfhighlight=/O,
  urlcolor=webbrown,
  linkcolor=RoyalBlue,
  filecolor=RoyalBlue,
  citecolor=webgreen,
  pdftitle={Physical limits to stochastic classifiers},
  pdfauthor={Timur Aslyamov and Massimiliano Esposito},
  pdfcreator={pdfLaTeX},
  pdfproducer={LaTeX REVTeX}
}

\usepackage[capitalize]{cleveref}

\newcommand{\llangle}{\langle\!\langle}
\newcommand{\rrangle}{\rangle\!\rangle}

\makeatletter
\def\maketag@@@#1{\hbox{\m@th\normalfont\normalsize#1}}
\makeatother

\newcommand{\me}{\mathrm{e}}

\newcommand{\E}{\mathbb{E}}

\newcommand{\dd}{\mathrm{d}}
\newcommand{\src}{\operatorname{s}}
\newcommand{\tgt}{\operatorname{t}}

\begin{document}
\title{Physical limits to stochastic classifiers}

\author{Timur Aslyamov}
\email{timur.aslyamov@uni.lu}
\affiliation{Complex Systems and Statistical Mechanics, Department of Physics and Materials Science, University of Luxembourg, 30 Avenue des Hauts-Fourneaux, L-4362 Esch-sur-Alzette, Luxembourg}

\author{Massimiliano Esposito}
\email{massimiliano.esposito@uni.lu}
\affiliation{Complex Systems and Statistical Mechanics, Department of Physics and Materials Science, University of Luxembourg, 30 Avenue des Hauts-Fourneaux, L-4362 Esch-sur-Alzette, Luxembourg}

\date{\today}

\begin{abstract}
Stochastic physical computers use random transitions between states to process inputs. 
We quantify their ability to solve distinct tasks as the control parameters of a device vary.
For Markov jump devices, we bound the number of classification tasks attainable at a prescribed output margin, with each task defined as a binary labeling of fixed inputs.
Combining geometric packing with fluctuation-response inequalities, we bound this number in terms of the control range and dimension, rate sensitivities, output fluctuations, and entropy production. 
For state and net-current observables, when controls change kinetics without changing thermodynamic forces and energies, the number of tasks is shaped by entropy production and output fluctuations. 
If the product of their maximum values over the allowed control settings and inputs is too small, the device can perform at most one task. An exactly solvable two-state switch illustrates how sufficient driving and observation time enable barrier tuning to select copying or inversion.
\end{abstract}

\maketitle

\textit{Introduction---}Programmable electronics must perform different operations with the same hardware: a signal that activates a circuit in one task may need to suppress it in another.
When charge or magnetization fluctuates, computation itself becomes stochastic, as in noisy electronic logic and networks of probabilistic bits~\cite{freitas2021stochastic,camsari2017stochastic}.
Such elements have been assembled into physical devices performing probabilistic computation~\cite{borders2019integer}, while physical learning provides methods to adjust device controls toward desired outputs~\cite{wright2022deep,momeni2025training}.
Their usefulness therefore depends on the number $N_{\rm tasks}$ of tasks accessible by retuning the same device.

Cells provide a natural example of this adaptability. They reuse the same metabolic network for different physiological demands by adjusting enzyme concentrations~\cite{zhu2023stringent}.
This makes metabolism an example of kinetic programming (control): enzymes change reaction speeds without changing chemical equilibrium.
Outside equilibrium, their activities can redistribute reaction fluxes and reshape metabolite concentrations~\cite{dai2018thermodynamic}.
This computational flexibility has been demonstrated in metabolic circuits programmed for different classifications by tuning enzyme levels~\cite{pandi2019metabolic} and in enzymatic neural networks with
adjustable nonlinear responses~\cite{okumura2022nonlinear}.
DNA assemblies also classify inputs through nucleation kinetics~\cite{evans2024pattern}.
Thus, kinetic regulation can give the same network different functions, but only while it is driven away from equilibrium or relaxing toward it.

Stochastic thermodynamics provides a natural framework for this research direction.
It describes fluctuating trajectories together with their work, heat, and entropy production, allowing computational outputs and energetic costs to be studied within the same dynamics~\cite{seifert2012stochastic}.
This approach has established connections between dissipation and learning~\cite{goldt2017stochastic}, the optimization of individual computations~\cite{lyu2026optimal,whitelam2025training,whitelam2025generative,whitelam2026evolutionary}, and the speed and efficiency of
thermodynamic computers~\cite{whitelam2026freedom}.
Related results constrain the errors of physical classifiers~\cite{hasegawa2026classifiers} and establish
speed--accuracy relations for diffusion models~\cite{ikeda2025speed}.

To estimate $N_{\rm tasks}$, we need to know how retuning the control parameters changes the mean output of a device. Response theory relates these changes to microscopic dynamics~\cite{baiesi2009fluctuations,seifert2010fluctuation,zheng2025spatial} and already constrains which computational functions a device can realize.
In steady state, varying a single transition rate while keeping the others fixed produces monotonic state occupations and local currents; therefore, an input acting monotonically through that rate cannot generate
a nonmonotonic classification~\cite{aslyamov2024nonequilibrium,aslyamov2024general,floyd2025local,floyd2025limits}.
The magnitude of steady-state response is further constrained by thermodynamic forces, network size, and
topology~\cite{owen2020universal,owen2023size,gabriela2023topologically,fancher2026topological,bao2024nonequilibrium}.
Exact nonequilibrium fluctuation-response relations (FRRs) establish a direct connection between fluctuations and control responses, expressing covariances in terms of responses to microscopic rate
perturbations~\cite{aslyamov2025fluctuation,ptaszynski2026frr,ptaszynski2026frr-mix}.
Dynamical FRRs extend this connection to transient evolution and time-dependent rates~\cite{aslyamov2026dfrr}, with related developments in the macroscopic limit~\cite{aslyamov2026macroscopic}, for overdamped and underdamped Langevin dynamics~\cite{chun2026fluctuation,vu2026exact}, and for non-Markovian jump processes~\cite{zheng2026memory}.
These identities yield response inequalities involving fluctuations and trajectory Fisher information~\cite{dechant2020fluctuation,zheng2024information,aslyamov2026dfrr}, alongside bounds governed by dynamical activity~\cite{liu2025dynamical} and entropy production~\cite{ptaszynski2024dissipation,aslyamov2025fluctuation,kwon2025fluctuation}.

In this Letter, we connect fluctuations and dissipation to the number of tasks that one stochastic device can realize within a prescribed control range and output margin.
Using fluctuation-response inequalities, we derive bounds that apply also at finite observation times.
For kinetic control of state and net-current observables, a sufficiently small product of the maximum output variance and maximum entropy production excludes the coexistence of two distinct attainable tasks.
General controls admit further bounds involving Fisher information and dynamical activity.
An exactly solvable two-state switch illustrates the restriction: driving enables barrier tuning to select copying or inversion, but the required margin becomes attainable only above a finite driving strength at a given observation time.

\textit{Physical binary classifier---}Binary classification assigns each input to one of two output classes;
such tasks also serve as building blocks for classification into more
classes~\cite{dietterich1995solving}.
For a physical realization, we measure an \textit{observable} $\mathcal Q$ during a run of duration $T$.
Its ensemble average is the \textit{mean output},
\begin{align}
    q(\bm\theta,\bm x)\equiv\E_{\bm\theta,\bm x}[\mathcal Q]\,,
    \label{eq:readout}
\end{align}
where $\bm x$ lists the input components and $\bm\theta=(\theta_1,\dots,\theta_{N_p})^\intercal$ specifies a
\textit{control setting}, a choice of values for all control parameters.
Throughout the main text, we assume that the observable $\mathcal{Q}$ does not depend explicitly on the control parameters $\bm \theta$; see its general expression in \cref{em:response-bounds}.
For example, observables include charge occupation, occupation time (time spent in a state),
and accumulated current (net transported charge) in quantum dots and mesoscopic
conductors~\cite{esposito2009nonequilibrium}.
The control parameters range over a convex domain $\Theta$ of finite radius $\Delta\theta$ about a
reference setting $\bm\theta_{\rm ref}$:
\begin{align}
    \sup_{\bm\theta\in\Theta}
      \|\bm\theta-\bm\theta_{\rm ref}\|_2=\Delta\theta\,,
    \label{eq:parameter-range}
\end{align}
where $\|\cdot\|_2$ is the Euclidean norm in dimensionless control coordinates.


As in supervised learning, controls are chosen to match desired
outputs~\cite{lecun2015deep,goodfellow2016deep}, without prescribing individual
trajectories. A \textit{task} assigns
binary \textit{labels} $y^\mu\in\{-1,+1\}$ to fixed \textit{inputs} $X=\{\bm x^\mu\}_{\mu=1}^{N_{\rm data}}$, giving the dataset
$\mathcal D=\{(\bm x^\mu,y^\mu)\}_{\mu=1}^{N_{\rm data}}$.
A task is \textit{attainable} if some control setting $\bm\theta$ in the
allowed range $\Theta$ satisfies
\begin{align}
    (q(\bm\theta,\bm x^\mu)-\eta) y^\mu \ge\gamma>0
    \qquad\text{for every }\mu\,,
    \label{eq:physical-classifier-margin}
\end{align}
where the \textit{output margin} $\gamma$ is the minimum required distance
from the fixed decision \textit{threshold} $\eta$: $q\ge\eta+\gamma$ for positive
labels and $q\le\eta-\gamma$ for negative labels. Each task must be solved
using one control setting shared by all inputs.

\begin{figure}[!t]
 \centering
 \includegraphics[width=\columnwidth]{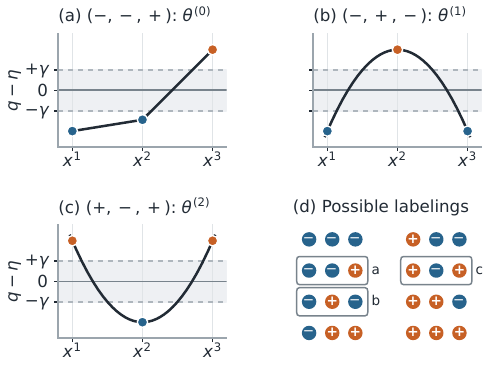}
 \caption{Tasks on the same three inputs $x^1<x^2<x^3$.
 Panels (a)--(c) illustrate schematic mean outputs at different control
 settings: (a) positive only at $x^3$; (b) positive only at $x^2$;
 (c) positive at both outer inputs $x^1$ and $x^3$ and negative at $x^2$.
 The shaded band is $|q-\eta|<\gamma$; all marked means meet the margin.
 Panel (d) lists all eight labelings and highlights these examples.
 The attainable number $N_{\text{tasks}}\le2^3$ depends on the device
 response and allowed controls.}
 \label{fig:fixed-architecture-tasks}
\end{figure}

With the inputs, threshold, and margin fixed, we define $Y$ as the set of all
attainable tasks (labelings). Their number is
\begin{align}
    N_{\text{tasks}}\equiv|Y|\le2^{N_{\rm data}}\,.
    \label{eq:N-task-def}
\end{align}
If no labeling meets the margin, $N_{\text{tasks}}=0$.
Figure~\ref{fig:fixed-architecture-tasks} illustrates the eight possible
tasks on three fixed inputs.
Suppose a device solves the task (a). Can retuning its control
parameters realize the nonmonotonic labeling (b) or (c)?
If $N_{\text{tasks}}=1$, no task other than the initial one is attainable.
\Cref{eq:N-task-def} gives a purely combinatorial upper bound. 
We seek tighter bounds shaped by the microscopic dynamics of the device.

\textit{Main results for kinetic control---}The most compact and physically insightful form of our results can be obtained for kinetic control, when $\bm \theta$ changes only the kinetics of transitions between the states of the system, but not the energies or thermodynamic forces. In biochemistry, such control corresponds to enzyme concentrations that catalyze metabolic reactions. In nanoelectronics, the kinetics is controlled by tunnel barriers of quantum dots and gate voltages for CMOS devices. 

For state and net-current observables under kinetic control, our results can be expressed in terms of the variance
\begin{align}
    \label{eq:var-def}
    \llangle\mathcal{Q}\rrangle \equiv \mathbb E_{\bm \theta, \bm x} [(\mathcal Q - q(\bm \theta, \bm x))^2]\,,
\end{align}
the physical entropy production $\Sigma(\bm \theta, \bm x)$, and the model-dependent transition rate sensitivity $U_B$. The latter encodes an explicit dependence of the model on the kinetic control $\bm \theta$, and one does not need to solve the dynamics to find it. Direct control of one dimensionless kinetic barrier gives $U_B=1$. The same holds for enzyme rates proportional to concentration $c$ when $\theta=\ln(c/c_{\rm ref})$.
Our first main result is a no-relabeling condition:
\begin{align}
    \llangle \mathcal Q \rrangle_{\max}\,\Sigma_{\max}<2\Big(\frac{\gamma}{\Delta\theta U_B}\Big)^2
    ~~\text{implies}~~N_{\rm tasks}\leq1\,,
    \label{eq:norelab-intro}
\end{align}
where $\Sigma_{\max} = \sup_{\bm \theta,\bm x} \Sigma$ and $\llangle \mathcal Q \rrangle_{\max} = \sup_{\bm \theta,\bm x} \llangle \mathcal Q \rrangle$ are the suprema over the convex domain $\bm \theta \in \Theta$ and the discrete list of inputs $\bm x \in X$.

For a device that remains at equilibrium, $\Sigma_{\max}^{\rm eq}=0$, and kinetic retuning cannot make new tasks attainable. 
Reducing only the output variance to zero yields $N_{\rm tasks}\leq1$.
Thus, under kinetic control, $N_{\rm tasks} > 1$ requires a sufficiently large product of fluctuations and entropy production. If this product is too small for all allowed controls and inputs, the device can realize at most one task.

As the second main result, we derive an upper bound
\begin{align}
    N_{\rm tasks} \leq \left\lfloor \Big(1 + \frac{\Delta \theta U_B}{\sqrt{2}\gamma}\sqrt{\llangle \mathcal{Q} \rrangle_{\max} \,\Sigma_{\max}}\Big)^{N_p}\right\rfloor\,,
    \label{eq:N-bound-intro}
\end{align}
where the floor takes the integer part. The practical potential of this bound is clear in a setup with $N_{\rm data}\gg N_p$, where \cref{eq:N-bound-intro} could be much tighter than the combinatorial \cref{eq:N-task-def}. 
In \cref{em:classification-capacity}, we relate our bounds to the Vapnik--Chervonenkis dimension and the results of Ref.~\cite{floyd2025limits}.

Next, we derive the no-relabeling condition and the upper bound on $N_{\rm tasks}$ for arbitrary control $\bm \theta \in \Theta$. We start with geometric constraints without any physical specifications. Then we bring in stochastic thermodynamical insight, using the control response of the mean output as the bridge.

\textit{Geometrical arguments---}Different tasks must assign opposite labels to at least one input $\bm x^\mu$, so their mean outputs must lie on opposite sides of the threshold, each at least $\gamma$ away [\cref{eq:physical-classifier-margin}].
Thus, their control settings $\bm\theta$ and $\bm\theta'$ must satisfy
\begin{align}
    |q(\bm\theta',\bm x^\mu)-q(\bm\theta,\bm x^\mu)|\ge2\gamma\,.
    \label{eq:different-tasks}
\end{align}
Moreover, using 
$\delta\bm\theta\equiv\bm\theta'-\bm\theta$ and $\bm\theta_u\equiv\bm\theta+u\delta\bm\theta$, we can represent the difference in \cref{eq:different-tasks} as the integral
\begin{align}
 q(\bm\theta',\bm x^\mu)-q(\bm\theta,\bm x^\mu)
 =\int_0^1\nabla_{\bm\theta}q(\bm\theta_u,\bm x^\mu)^\intercal
       \delta\bm\theta\,\dd u\,,
 \label{eq:segment-identity}
\end{align}
where $\bm\theta_u \in \Theta$ for every $u\in[0,1]$ by convexity.
The control response of the mean output is the vector $\nabla_{\bm\theta}q$, whose component $\partial_{\theta_i}q$ gives the response to an infinitesimal change in control parameter $\theta_i$.
The difference can be bounded by
\begin{subequations}
\begin{align}
    |q(\bm\theta',\bm x^\mu)&-q(\bm\theta,\bm x^\mu)|
       \le K(\bm x^\mu)\|\bm\theta'-\bm\theta\|_2\,,
    \label{eq:input-lipschitz}\\
    K(\bm x)&\equiv\sup_{\bm\theta\in\Theta}
        \|\nabla_{\bm\theta}q(\bm\theta,\bm x)\|_2\,.
    \label{eq:input-response-factor}
\end{align}
\end{subequations}
Here $K(\bm x)$ bounds the magnitude of the control response uniformly over $\Theta$ for $\bm x \in X$.
For a fixed input $\bm x \in X$, changing the label at margin $\gamma$ requires
\begin{align}
 2\gamma\le K(\bm x)\|\bm\theta'-\bm\theta\|_2\le2\Delta\theta K(\bm x).
 \label{eq:input-separation}
\end{align}
Counting such inputs $m_\gamma\equiv\#\{\bm x\in X:\Delta\theta K(\bm x)\ge\gamma\}$ gives a bound on the number of attainable tasks:
\begin{align}
    N_{\text{tasks}}&\le2^{m_\gamma}\,,
    \label{eq:input-task-count}
\end{align}
which is tighter or equal than the combinatorial bound \eqref{eq:N-task-def}.

For two tasks satisfying \cref{eq:physical-classifier-margin}, the condition \eqref{eq:input-separation} must be satisfied for every input whose label differs between them.
This results in $\gamma\le\Delta\theta K_{\max}$, where we use the uniform control-response bound $K_{\max} \equiv \sup_{\bm x\in X}K(\bm x)$, which bounds the response over all allowed controls and inputs.
\phantomsection
\label{thm:no-learning}
Inverting this statement, we formulate a \textit{no-relabeling condition}:
\begin{align}
    \Delta\theta K_{\max}<\gamma~~\text{implies}~~N_{\rm tasks} \leq 1\,.
    \label{eq:no-learning-geometric}
\end{align}
If an initial setting realizes a labeling at margin $\gamma$, no allowed control change can realize a different labeling of the same inputs at that margin.

\begin{figure}[t]
 \centering
 \includegraphics[width=0.6\linewidth]{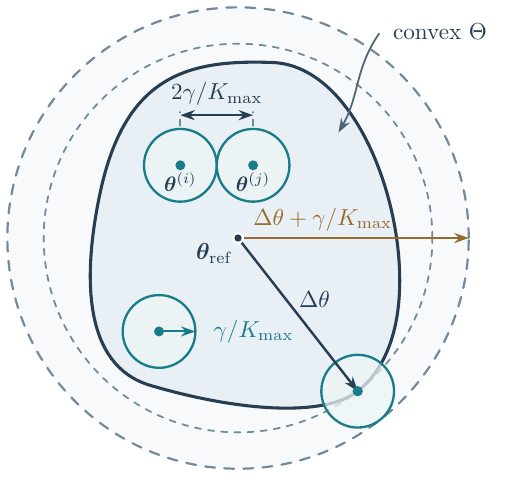}
 \caption{Geometric packing of attainable tasks in control space.
 Each small ball is centered at a setting realizing a different task.
 The centers are at least $2\gamma/K_{\max}$ apart, so the open balls of radius
 $\gamma/K_{\max}$ are disjoint. Comparing their total volume with the enclosing
 ball gives the task-count bound.}
 \label{fig:task-ball-packing}
\end{figure}

Now we derive a geometric task-count bound. For $K_{\max}>0$, one can place an open ball of radius
$\gamma/K_{\max}$ around one setting realizing each attainable task. 
These balls are disjoint and fit inside a ball of radius $\Delta\theta+\gamma/K_{\max}$ centered at $\bm\theta_{\rm ref}$. 
For illustration, see the sketch in Fig.~\ref{fig:task-ball-packing}.
Comparing their volumes~\cite[Sec.~4.2]{vershynin2018high} gives
\begin{align}
    N_{\text{tasks}}\left(\frac{\gamma}{K_{\max}}\right)^{N_p}
    \le\left(\Delta\theta+\frac{\gamma}{K_{\max}}\right)^{N_p}\,.
    \label{eq:volume-comparison}
\end{align}
Rearranging \cref{eq:volume-comparison}, we find the bound
\begin{align}
    N_{\text{tasks}}\le \left\lfloor\left(1+\frac{\Delta\theta K_{\max}}{\gamma}\right)^{N_p}
       \right\rfloor\,.
    \label{eq:task-bound-geometric}
\end{align}
If $K_{\max}=0$ or
$\Delta\theta=0$, the mean output $q$ cannot change with the controls and at
most one task is attainable.

\textit{Stochastic thermodynamics of a device---}Starting from here, we bring the device's physics into consideration.
We specify a finite graph $\mathcal{G}$ of system states (nodes) and physical transition channels (edges) between them.
The state probabilities $\bm p = (p_1, p_2, \dots)^\intercal$ obey the
master equation of a possibly time-dependent Markov jump process,
\begin{align}
    \partial_t\bm p(t)=W(t)\bm p(t),\qquad
    \bm1^\intercal\bm p=1,\quad\bm1^\intercal W=0,
    \label{eq:master}
\end{align}
where $W(t;\bm\theta,\bm x)$ is the rate matrix; we suppress input and control arguments at a fixed setting. 
The rates may depend explicitly on time, while the control parameters $\bm\theta$ remain constant during each run.
The preparation $\bm p(0)$, the observation time $T$, the graph $\mathcal{G}$, and the form of input dependence remain fixed as controls vary.

In the graph $\mathcal{G}$, a directed edge $e$ denotes a transition from its source $\src(e)$ to its target $\tgt(e)$; $-e$ denotes the reverse edge. Its incidence vector has components $(\bm\Delta_e)_n=\delta_{n,\tgt(e)}-\delta_{n,\src(e)}$, so $\bm\Delta_e\cdot\bm Q=Q_{\tgt(e)}-Q_{\src(e)}$. 
The off-diagonal entry $W_{nm}$ sums all transition rates from $m$ to $n$, and the diagonal entries enforce probability conservation.  
For each transition, the directed flux, net current, and activity are
$j_e\equiv W_ep_{\src(e)}$, $J_e\equiv j_e-j_{-e}$, and $a_e\equiv j_e+j_{-e}$, respectively. The latter is also known as traffic and frenesy. 
The sum $\sum_{e>0}$ includes one orientation per reversible channel, whereas $\sum_e$ includes both orientations.
In units with $k_B=1$, the physical entropy production and total activity are
\begin{align}
    \Sigma\equiv\int_0^T\dd t\sum_{e>0}J_e\ln\frac{j_e}{j_{-e}}\,,
    \qquad A\equiv\int_0^T\dd t\sum_{e>0}a_e\,.
    \label{eq:entropy-activity-def}
\end{align}
In addition, we define the Fisher information rate matrix
\begin{align}
    \dot I(t)\equiv\sum_e j_e(t)\bm U_e(t)\bm U_e(t)^\intercal\,,\qquad \bm U_e\equiv\nabla_{\bm\theta}\ln W_e\,,
    \label{eq:information-matrix}
\end{align}
which combines log-rate sensitivities $\bm U_e$ in a bilinear form. It defines the Fisher information matrix as $I\equiv \int_0^T\dd t \dot I(t)$.

The key element connecting the response with the thermodynamic and information observables is the variance~\eqref{eq:var-def}. However, not all of the total variance is controlled by responses~\cite{aslyamov2026dfrr}. There is a part induced by the noise from the preparation ensemble, with $\bm p(0)$ independent of the control and input parameters. To find the contribution of dynamical variance, we subtract the initial variability from the total variance in \cref{eq:var-def}:
\begin{align}
\label{eq:V-def}
    V(\bm\theta,\bm x) &\equiv \llangle \mathcal Q \rrangle_{\bm\theta,\bm x} - V_0 \leq \llangle \mathcal Q \rrangle_{\bm\theta,\bm x}\,.
\end{align}
Here, the initial variability $V_0 \equiv \sum_{n}p_n(0)\delta Q^2_n(0;\bm\theta,\bm x)$ is defined in terms of the centered conditional means $\delta Q_n(0) \equiv \mathbb{E}(\mathcal Q\vert n_0 = n) - q(\bm\theta,\bm x)$. The latter is a dynamical extension of the excess observable~\cite{aslyamov2025excess}. 



\textit{Results for arbitrary control---}For detailed derivations of results from this section, see \cref{em:response-bounds}. 
We start with the Fisher information bound.
The fluctuation-response inequality (FRI), corrected for initial
variability~\cite{aslyamov2026dfrr}, bounds the control-response magnitude uniformly over controls and inputs as
\begin{align}
    K_{\max}\le\Gamma_{\rm F}\equiv\sqrt{V_{\max}I_{\max}},
    \label{eq:gamma-Fisher}
\end{align}
where, introducing $\|I\|_{\rm op}$ the largest eigenvalue of $I$,
\begin{align}
    V_{\max}&=\sup_{\bm\theta\in\Theta,\,\bm x\in X}V,
       &I_{\max}&=\sup_{\bm\theta\in\Theta,\,\bm x\in X}\|I\|_{\rm op}.
    \label{eq:uniform-information}
\end{align}

For the thermodynamic bounds, we use a mathematically exact and physically motivated parameterization that enforces thermodynamic consistency:
\begin{align}
    W_{\pm e}(t;\bm\theta,\bm x)
    =\exp[-B_e(t;\bm\theta,\bm x)\pm S_e(t;\bm\theta,\bm x)/2],
    \label{eq:physical-classifier-rates}
\end{align}
where $B_e\equiv-\tfrac12\ln(W_eW_{-e})=B_{-e}$ is a kinetic (barrier) term, and
$S_e\equiv\ln(W_e/W_{-e})=-S_{-e}$ is a thermodynamic term (the entropy
transferred to the reservoirs along $e$). 
The latter has a further decomposition $S_e=F_e-\bm\Delta_e\cdot\bm\Phi$,
where $\bm\Phi$ is the dimensionless state potential and $F_e=-F_{-e}$ is the nonconservative force on edge $e$~\cite{rao2018conservation,falasco2025macroscopic}.

To proceed, let $\Sigma_{\max}\equiv\sup_{\bm\theta,\bm x}\Sigma$ and
$A_{\max}\equiv\sup_{\bm\theta,\bm x}A$, and define
the uniform susceptibilities
\begin{align}
    U_B&\equiv\sup_{e>0,t,\bm\theta,\bm x}\|\nabla_{\bm\theta}B_e\|_2\,,
       &U_S&\equiv\sup_{e>0,t,\bm\theta,\bm x}\|\nabla_{\bm\theta}S_e\|_2\,,
    \label{eq:barrier-force-susceptibilities}
\end{align}
which correspond to the highest control sensitivities of the kinetic and energetic parameters.
Restricting ourselves to state and net current observables (the transition part of $\mathcal Q$ must change sign under reversal) [see \cref{em:response-bounds}], we get the thermodynamic bound:
\begin{align}
    K_{\max}\le\Gamma_{\rm th}\equiv\sqrt{V_{\max}}
       \left[U_B\sqrt{\frac{\Sigma_{\max}}2}
             +\frac{U_S}{2}\sqrt{A_{\max}}\right]\,.
    \label{eq:lipschitz-thermo}
\end{align}

Now we combine the geometrical arguments from \cref{eq:no-learning-geometric,eq:task-bound-geometric} with the bounds in \cref{eq:lipschitz-thermo,eq:gamma-Fisher}.
For observables satisfying both bounds, we find 
the sufficient no-relabeling condition
\begin{align}
    \Delta\theta\min\{\Gamma_{\rm F},\Gamma_{\rm th}\}<\gamma~~\text{implies}~~N_{\rm tasks} \leq 1\,.
    \label{eq:norelab}
\end{align}
The physical version of the bound
\eqref{eq:task-bound-geometric} reads
\begin{align}
    N_{\text{tasks}}\le
       \left\lfloor\left(1+\frac{\Delta\theta}{\gamma}
        \min\{\Gamma_{\rm F},\Gamma_{\rm th}\}\right)^{N_p}
       \right\rfloor\,.
    \label{eq:N-bound}
\end{align}
\cref{eq:norelab,eq:N-bound} generalize \cref{eq:norelab-intro,eq:N-bound-intro} which are recovered from them using $V_{\max}\leq\llangle\mathcal Q\rrangle_{\max}$ and assuming kinetic control ($U_S = 0$). 

\begin{figure}
 \centering
 \includegraphics[width=\columnwidth]{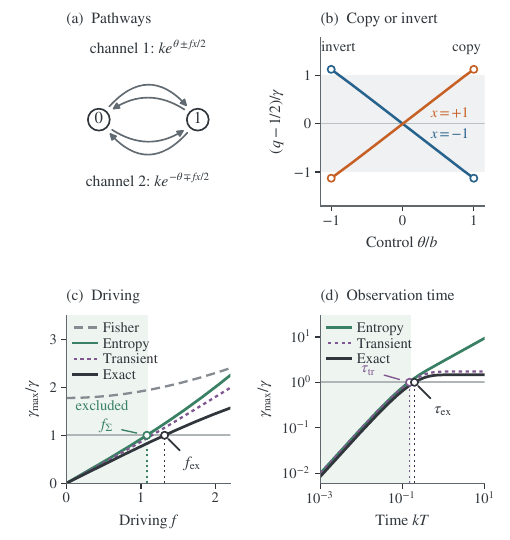}
 \caption{Programming the two-state switch, with $x=\pm1$, uniform
 preparation, $\eta=1/2$, $\gamma=0.10$, $b=0.5$, $f=1.5$, and $kT=0.25$
 unless varied. (a) The control changes the two channel barriers oppositely.
 (b) Centered mean outputs divided by $\gamma$ for $x=-1$ (blue) and $+1$ (orange);
 the endpoints realize inversion and copying.
 (c,d) Normalized maximum attainable margin
 $\gamma_{\max}^{\rm exact}/\gamma$ (black) and its thermodynamic,
 transient, and Fisher upper bounds versus driving and observation time.
 The transient curve uses $b\bar\Gamma_{\rm tr}$.
 Shading marks task exclusion by the thermodynamic bound in (c) and the
 transient bound in (d); the horizontal line marks the required margin.}
 \label{fig:programmable-switch}
\end{figure}

\textit{Application: a programmable two-state switch---}A sensor signal may need to turn the same device on in one task and off in another.
Copying preserves the signal's high or low state; inversion reverses it.
Here, using a two-state switch as an example, we find the driving force and observation time required for copying and inversion tasks.
A possible realization is a quantum dot with two charge states, contacts whose rates are gate-tunable, and a charge sensor for readout~\cite{maradan2014thermometry}.
The setup consists of two states $0$ and $1$ connected by two reversible channels. The input $x\in\{-1,+1\}$ biases them in opposite
directions, while a shared control $\theta\in[-b,b]$ lowers one barrier
and raises the other:
\begin{align}
    W_{\pm 1}(\theta,x)&=k\me^{\theta\pm fx/2}\,,&
    W_{\pm 2}(\theta,x)&=k\me^{-\theta\mp fx/2}\,,
    \label{eq:switch-rates}
\end{align}
where $k>0$ sets the rate scale, $f\ge0$ is the driving amplitude biasing
the channels, and $+$ denotes $0\to1$. 
The nonconservative cycle force is $\ln\frac{W_{+1}W_{-2}}{W_{-1}W_{+2}}=2fx$.
We use a kinetic control that changes the competition between pathways
without changing their driving.
Thus, the control parameter has $\partial_\theta B_1=-1$,
$\partial_\theta B_2=1$, and $U_B=1$, $U_S=0$.

The terminal observable $\mathcal Q=\mathbf1_{\{n_T=1\}}$ equals one
if the state at time $T$ is $1$, and zero otherwise. The mean output is $q(\theta,x) = p_1(T)$. With the uniform
preparation $p_0(0)=p_1(0)=1/2$ we find
\begin{align}
    q(\theta,x)  =\frac12+
       \frac{\tanh\theta\,\tanh(fx/2)}2(1-\me^{-\lambda(\theta) T})\,,
    \label{eq:switch-mean}
\end{align}
which suggests the natural threshold $\eta=1/2$. 
Here $\lambda=4k\cosh\theta\cosh(f/2)$ is the relaxation scale. 
This shows that $\theta > 0$ favors copying the input sign; $\theta < 0$ favors inversion~[\cref{fig:programmable-switch}(a,b)].
At $f=0$, the uniform preparation is stationary for every control, so barrier changes cannot alter the occupation.
Thus, driving allows the barrier to control the mean output.

End Matter~\ref{em:switch-bounds} uses the exact fluctuations, entropy production, and activity to establish the
uniform estimates
\begin{subequations}
\label{eq:switch-uniform-budgets}
\begin{align}
    V_{\max}&\le\bar V=\frac{1-\me^{-2 T\Lambda}}4\,,
      \\
    \Sigma_{\max}&\le\bar\Sigma=2kTf\cosh b\sinh(f/2)\,,
      \\
    A_{\max}&\le\bar A= T\Lambda/2\,, \qquad I_{\max}\le\bar A\,,
\end{align}
\end{subequations}
where $\Lambda\equiv\lambda(b)$. Here the Fisher information equals activity
because $U_e^2 = 1$ in \cref{eq:information-matrix}.
These estimates give upper bounds on $K_{\max}$, $\bar\Gamma_{\rm F}=\sqrt{\bar V\bar A}$ and $\bar\Gamma_{\rm th}=\sqrt{\bar V\bar\Sigma/2}$.
For this purely kinetic control, the thermodynamic bound depends on entropy production and is labeled ``Entropy'' in the figures.

Retaining the relaxation dynamics in
\cref{em:eq:gamma-transient,em:eq:switch-transient} gives
$K_{\max}\le\Gamma_{\rm tr}\le\bar\Gamma_{\rm tr}(T)$, where
\begin{align}
 \bar\Gamma_{\rm tr}(T)
 &\equiv(1-\me^{- T\Lambda})
   \min\!\left\{\sqrt{\frac{f\tanh(f/2)}8},\frac12\right\}\,,
 \label{eq:switch-gamma-transient}
\end{align}
At fixed nonzero driving, $\bar\Gamma_{\rm tr}(T)$ saturates at long
times, whereas $\bar\Gamma_{\rm F}$ and $\bar\Gamma_{\rm th}$ grow as $\sqrt T$.
The activity term also bounds $\bar\Gamma_{\rm tr}(T)$ by $1/2$
as driving increases. 

Since $q(0,x)=1/2$ and $\Delta\theta=b$, the no-relabeling condition
$b K_{\max}<\gamma$ prevents either task
from reaching the required margin. Thus, to solve copying and inversion tasks, the required margin must satisfy the theoretical upper bound
\begin{align}
    \gamma \leq \gamma_{\max}^{\rm theory} \equiv b \min \{\bar\Gamma_{\rm F},\bar\Gamma_{\rm th}, \bar\Gamma_{\rm tr}(T)\}\,.
    \label{eq:max-gamma-theory}
\end{align}
We test these bounds against the exact maximum attainable margin. For either input, the displacement $|q(\theta,x)-\eta|$ is
nondecreasing in $|\theta|$ and reaches its maximum at $\theta=\pm b$.
Maximizing the smaller margin of the two inputs gives the common maximum for copying and inversion:
\begin{align}
    \gamma_{\max}^{\rm exact}
       &=q(b,+1)-\frac12
        =\frac{\tanh b\,\tanh(f/2)}2(1-\me^{- T\Lambda})\,.
    \label{eq:max-gamma-exact}
\end{align}
The two tasks are attainable precisely when $0<\gamma\le\gamma_{\max}^{\rm exact}$.
The symmetry $q(\theta,+1)+q(\theta,-1)=1$ excludes both constant
labelings, $\bm y=(+1,+1)$ and $\bm y=(-1,-1)$, so
$N_{\text{tasks}}$ is exactly two in this regime and zero otherwise.

\Cref{fig:programmable-switch}(c) shows that realizing $N_{\rm tasks}=2$ requires sufficiently strong driving:
a nonzero force alone is not enough.  
The thermodynamic bound captures this threshold. Its lower bound $f_\Sigma$ on the required driving is close to the
exact threshold $f_{\rm ex}$. 
\Cref{fig:programmable-switch}(d) shows that sufficient driving must also act long enough. 
The transient bound accounts for the fading influence of early jumps on the final state and also saturates. 
Writing the dimensionless observation time as $\tau=kT$, the transient lower bound $\tau_{\rm tr}$ on the
required duration is about $76\%$ of the exact minimum $\tau_{\rm ex}$
for the parameters in \cref{fig:programmable-switch}(d).
For $b,f,T>0$, the transient bound and the exact margin share the
relaxation factor $1-\me^{-T\Lambda}$, so their ratio is independent
of observation time. This ratio approaches one for weak driving and
small control range, although the attainable margin then vanishes.
In addition, \cref{fig:switch-margin-thresholds} shows how the
minimum force and time depend on the required margin.

\textit{Conclusions---}We have bounded how many tasks a stochastic device can realize by retuning its controls. 
For state and net-current observables under kinetic control, attaining $N_{\rm tasks} > 1$ at a fixed output margin
requires a sufficiently large product of the variance of the observable and entropy production.
If this product is too small across all allowed kinetics and inputs, the device
cannot support more than one task. 
In particular, barrier changes cannot reprogram a device that stays at equilibrium throughout its run, even when jumps remain frequent. 
The two-state switch shows how driving makes $N_{\rm tasks} > 1$ attainable, but only above a finite driving strength and with a sufficiently long
observation time.

\begin{acknowledgments}
T.A. and M.E. are funded by the Fonds National de la Recherche-FNR, Luxembourg: project ThermoElectroChem (C23/MS/18060819) and MacroFluctResp (C25/MS/19444229), respectively.
\end{acknowledgments}

\bibliographystyle{apsrev4-2}
\bibliography{references}

\clearpage
\section*{End Matter}
\appendix
\setcounter{secnumdepth}{2}
\renewcommand{\theHsection}{em.\Alph{section}}
\renewcommand{\theHequation}{em.\Alph{section}.\arabic{equation}}

\section{Response bounds from dynamical fluctuations}
\label{em:response-bounds}


\textit{Variance and response---}Adapting the response and variance formulas of Ref.~\cite{aslyamov2026dfrr} to include a terminal contribution, we derive the uniform Fisher-information, thermodynamic, and transient bounds on $K_{\max}$ used in the main text.

We define the observable through terminal, state, and jump contributions,
\begin{align}
 \mathcal Q[\omega_{0:T}]
 &\equiv g(n_T)+\int_0^T o_{n_t}\,\dd t+\sum_e\int_0^T v_e\,\dd k_e\,,
 \label{eq:observable}
\end{align}
where $\dd k_e(t)$ counts jumps along $e$. The weights $g(n_T;\bm\theta)$, $o_n(t;\bm\theta)$, and $v_e(t;\bm\theta)$ may depend on the controls; we suppress these arguments below.
Let $\mathcal Q_{t:T}$ denote the part of $\mathcal Q$ still to be accumulated after $t$, including the terminal term, and define
$Q_n(t)\equiv\E[\mathcal Q_{t:T}\mid n_t=n]$ as the conditional mean, which satisfies the backward equation:
\begin{align}
 -\partial_t Q_n&=o_n+\sum_{e:\,\src(e)=n}W_e\varphi_e,
 \qquad Q_n(T)=g(n).
 \label{em:eq:backward}
\end{align}
Here $\varphi_e(t)\equiv v_e(t)+\bm\Delta_e\cdot\bm Q(t)$ is the change in expected output caused by a jump,
including its immediate contribution. 

Next, we use Eq.~(11) from Ref.~\cite{aslyamov2026dfrr} to express the dynamical variance $V = \llangle\mathcal Q\rrangle-V_0$ [\cref{eq:V-def}] as
\begin{align}
 V&=\int_0^T\nu_T(t)\,\dd t,
 &\nu_T(t)&\equiv\sum_e j_e(t)\varphi_e(t)^2\,,
 \label{em:eq:variance}
\end{align}
where $\nu_T(t)$ is the variance density.
In addition, Eq.~(9) in Ref.~\cite{aslyamov2026dfrr} gives the control response of the mean output as
\begin{align}
 \nabla_{\bm\theta}q = \bm D + \int_0^T\dd t\sum_e j_e\varphi_e\bm U_e\,,
 \label{em:eq:response}
\end{align}
where $\bm U_e = \nabla_{\bm\theta}\ln W_e$ also defines the Fisher information matrix; see \cref{eq:information-matrix}. 
Here, $\bm D\equiv\E[\nabla_{\bm\theta}\mathcal Q]$ encodes the explicit dependence of the observable on $\bm \theta$. For the observable in \cref{eq:observable}, it is
\begin{align}
 \bm D&=\sum_n p_n(T)\nabla_{\bm\theta}g(n) + \int_0^T\dd t\Big[
   \sum_n p_n\nabla_{\bm\theta}o_n
   +\sum_e j_e\nabla_{\bm\theta}v_e\Big].
 \label{em:eq:direct-response}
\end{align}

\textit{Fluctuation-response inequality---}For any fixed control direction $\bm u$, using \cref{em:eq:response} and applying the Cauchy--Schwarz inequality to the sum $\sum_e$, we find
\begin{align}
 |\bm u^\intercal(\nabla_{\bm\theta}q-\bm D)|
 &\le\int_0^T\dd t\,
       \sqrt{\nu_T(t)\,\bm u^\intercal\dot I(t)\bm u}.
 \label{em:eq:transient-fisher}
\end{align}
Next, using the Cauchy--Schwarz inequality in time, we arrive at the initial-variability-corrected
FRI~\cite{aslyamov2026dfrr},
\begin{align}
 |\bm u^\intercal(\nabla_{\bm\theta}q-\bm D)|^2
 &\le\left(\int_0^T\nu_T\,\dd t\right)
       \left(\int_0^T\bm u^\intercal\dot I\bm u\,\dd t\right)
=V\,\bm u^\intercal I\bm u.
 \label{em:eq:fisher}
\end{align}
Here, removing initial variability in $V$, \cref{em:eq:fisher} recovers the fluctuation-response inequality [Eqs.~(12) and (15) in Ref.~\cite{dechant2020fluctuation}].
To optimize over unit directions, we use
$\sup_{\|\bm u\|_2=1}\bm u^\intercal I\bm u=\|I\|_{\rm op}$.
Applying the triangle inequality to \cref{em:eq:fisher} and using \cref{eq:uniform-information} gives
\begin{align}
\label{em:eq:FRI}
 K_{\max}
 &\le\sup_{\bm\theta\in\Theta,\,\bm x\in X}
   \left[\|\bm D\|_2+\sqrt{V\|I\|_{\rm op}}\right]\le D_{\max}+\Gamma_{\rm F}\,,
\end{align}
where $\Gamma_{\rm F}$ is defined in \cref{eq:gamma-Fisher}, and we introduce $D_{\max}\equiv\sup_{\bm\theta\in\Theta,\,\bm x\in X}\|\bm D\|_2<\infty$.
For control-independent weights, $\bm D=0$, and we recover
\cref{eq:gamma-Fisher}.

\textit{Transient bounds---}A thermodynamic refinement applies when $v_{-e}=-v_e$.
Then $\varphi_{-e}=-\varphi_e$ and
$\nu_T=\sum_{e>0}a_e\varphi_e^2$.
Substituting the rates from \eqref{eq:physical-classifier-rates} into \cref{em:eq:response} yields
\begin{align}
 \nabla_{\bm\theta}q-\bm D
 &=\int_0^T\dd t\sum_{e>0}\varphi_e
   \left[-J_e\nabla_{\bm\theta}B_e
         +\frac{a_e}{2}\nabla_{\bm\theta}S_e\right].
 \label{em:eq:paired-response}
\end{align}
After applying the Cauchy--Schwarz inequality to the sum $\sum_{e>0}$, we have
\begin{align}
 \|\nabla_{\bm\theta}q-\bm D\|_2
 &\le\int_0^T\dd t\,\sqrt{\nu_T(t)}
   \left[U_B\sqrt{\dot\Pi(t)}
         +\frac{U_S}{2}\sqrt{\dot A(t)}\right]\,,
 \label{em:eq:time-resolved-response}
\end{align}
where $\dot A(t)\equiv\sum_{e>0}a_e(t)$ is the activity rate and
$\dot\Pi\equiv\sum_{e>0}J_e^2/a_e$ is the pseudo-entropy production rate,
with zero contribution from channels with $a_e=0$.
The entropy-production rate obeys $\dot\Sigma\ge2\dot\Pi$.
Also, $|J_e|\le a_e$ implies $J_e^2/a_e\le a_e$, so
\begin{align}
 \dot\Pi(t)\le\min\!\left\{\frac{\dot\Sigma(t)}2,\dot A(t)\right\}.
 \label{em:eq:pseudo-entropy-activity}
\end{align}
Applying this inequality and taking the supremum over controls and inputs gives
$K_{\max}\le D_{\max}+\Gamma_{\rm tr}$, with the explicit definition
\begin{align}
 \Gamma_{\rm tr}
 &\equiv\sup_{\bm\theta\,,\bm x}
   \int_0^T\dd t\,\sqrt{\nu_T(t)}\nonumber\\
 &\quad\times\left[U_B\sqrt{\min\!\left\{\dot\Sigma(t)/2,\dot A(t)\right\}}
       +\frac{U_S}{2}\sqrt{\dot A(t)}\right]\,.
 \label{em:eq:gamma-transient}
\end{align}
The minimum is taken at each time and setting before integration and the
supremum. It strengthens the entropy-only transient estimate while retaining
its equilibrium limit for kinetic control. Saturation of the response with
time or driving additionally depends on the observable and its relaxation;
it does not follow from the minimum alone.

\textit{Integrated thermodynamic bound---}For antisymmetric jump weights, the Cauchy--Schwarz inequality in time gives
\begin{align}
 \int_0^T\sqrt{\nu_T(t)\dot\Sigma(t)}\,\dd t
 &\le\sqrt{V\Sigma}\,,
 &\int_0^T\sqrt{\nu_T(t)\dot A(t)}\,\dd t
 &\le\sqrt{VA}\,.
 \label{em:eq:time-cauchy-schwarz}
\end{align}
Bounding the minimum in \cref{em:eq:gamma-transient} by $\dot\Sigma/2$
and applying these inequalities separately to the two terms recovers
\begin{align}
 \|\nabla_{\bm\theta}q-\bm D\|_2
 &\le\sqrt V\left[U_B\sqrt{\Sigma/2}
                   +\frac{U_S}{2}\sqrt A\right]\,.
 \label{em:eq:thermodynamic}
\end{align}
For control-independent weights, $\bm D=0$, and we recover the thermodynamic bound $K_{\max} \leq \Gamma_{\rm th}$ from the main text.

\section{Relation to Vapnik--Chervonenkis dimension}
\label{em:classification-capacity}

Here we return to the control-independent state and net current observables, $\bm D = 0$ and $v_{-e} = -v_{e}$.
For the same fixed protocol, the capacity $C(\gamma)$ is the largest
number of inputs for which every binary labeling is attainable at
margin $\gamma$:
\begin{align}
 C(\gamma)\equiv\max_X\bigl\{|X|:\,
 N_{\rm tasks}(X,\gamma)=2^{|X|}\bigr\}.
 \label{em:eq:capacity-definition}
\end{align}
With the response bounds evaluated over all allowed inputs,
substituting $N_{\rm tasks}=2^{|X|}$ into the task-count bound gives
\begin{align}
 C(\gamma)
 \le\left\lfloor N_p\log_2\!\left(1+
       \frac{\Delta\theta\Gamma_\alpha}{\gamma}\right)\right\rfloor\,,\qquad \alpha \in \{\text{F, th, tr}\}\,.
 \label{em:eq:margin-capacity}
\end{align}
The Vapnik--Chervonenkis (VC) dimension $d_{\rm VC}$ asks the same
question without a prescribed positive margin~\cite[Chap.~6]{shalev2014understanding}.
Thus $C(\gamma)\le d_{\rm VC}$, but an upper bound on
$C(\gamma)$ does not generally upper-bound $d_{\rm VC}$.

For stationary occupations with the exponential input coupling
studied by Floyd \emph{et al.}~\cite{floyd2025limits}, their bound gives
\begin{align}
 C_{\rm ss}(\gamma)
 \le d_{\rm VC,ss}\le(2M+1)^d,
 \label{em:eq:stationary-capacity}
\end{align}
where $d$ is the number of input components and $M$ is the number
of edges affected by each input component. When both bounds apply
to the same stationary observable, neither is universally tighter.

\section{Bounds for the binary switch application}
\label{em:switch-bounds}

For the switch in \cref{eq:switch-rates}, write
$\rho_x=\tanh\theta\tanh(fx/2)$ and use
$\lambda=4k\cosh\theta\cosh(f/2)$.
The terminal observable has ${v_{\pm e}=0}$. Both edges
$e=1,2$ describe $0\to1$ transitions, so \cref{em:eq:backward} gives
$\varphi_{+e}(t)=\varphi(t)$ and $\varphi_{-e}(t)=-\varphi(t)$, where
\begin{align}
 \varphi(t)&\equiv Q_1(t)-Q_0(t)=\me^{-\lambda(T-t)}\,.
 \label{em:eq:switch-backward}
\end{align}
For $\bm p(0)=(1/2,1/2)^\intercal$, the initial variability is $V_0=\me^{-2\lambda T}/4$.
The exact mean output \eqref{eq:switch-mean} gives the total activity rate
\begin{align}
 a(t)=\frac\lambda2
       [1-\rho_x^2(1-\me^{-\lambda t})]\le\frac\lambda2.
 \label{em:eq:switch-activity}
\end{align}
Hence $A\le\lambda T/2\le T\Lambda/2$. Every directed log-rate
score has magnitude one, so $I=A$.
For the terminal indicator observable, we have $\llangle \mathcal Q \rrangle = q(1-q)$. Together with \cref{eq:V-def}, this gives
\begin{align}
 V=q(1-q)-\frac{\me^{-2\lambda T}}4
   \le\frac{1-\me^{-2 T\Lambda}}4=\bar V,
 \label{em:eq:switch-variance}
\end{align}
where we used $q(1-q) \leq 1/4$.

Taking the internal entropy of each switch state to be zero, the mean entropy production rate decomposes as~\cite[Sec.~II]{falasco2025macroscopic}
\begin{align}
\label{em:eq:switch-entropy}
\dot\Sigma&=\dot S_{\rm env}+\dot S \leq \dot S_{\rm env}\,,
\end{align}
where $\dot S_{\rm env}\equiv\sum_{e>0}J_eS_e=fx(J_1-J_2)$ is the mean entropy-flow rate into the environment and $S=-p_1\ln p_1-(1-p_1)\ln(1-p_1)$ is the Shannon entropy.
The inequality in \cref{em:eq:switch-entropy} holds because the occupation moves monotonically away from its initial value $p_1(0) = 1/2$, so
$\dot S=\dot p_1\ln[(1-p_1)/p_1]\leq0$.
Consequently,
\begin{align}
 \dot\Sigma \leq \dot S_{\rm env} &=\frac{\lambda f\tanh(f/2)}2
   [1-\tanh^2\theta(1-\me^{-\lambda t})]
 \nonumber\\
 &\le\frac{\lambda f\tanh(f/2)}2.
 \label{em:eq:switch-entropy-bound}
\end{align}
Integrating over time and using $\lambda\le\lambda(b)$, we arrive at
$\Sigma\le2kTf\cosh b\sinh(f/2)=\bar\Sigma$.
\Cref{em:eq:switch-activity,em:eq:switch-variance,em:eq:switch-entropy-bound} prove all bounds in \eqref{eq:switch-uniform-budgets}.

For this purely kinetic control, $U_B=1$, $U_S=0$, and
$\nu_T(t)=a(t)\varphi(t)^2$. Substitution into
\cref{em:eq:gamma-transient}, using
\cref{em:eq:switch-activity,em:eq:switch-entropy-bound}, bounds its integral by
\begin{align}
 &\int_0^T\dd t\,\varphi(t)
   \sqrt{a(t)\min\!\left\{\dot\Sigma(t)/2,a(t)\right\}}
   \nonumber\\
 &\quad\le(1-\me^{-\lambda T})
   \min\!\left\{\sqrt{\frac{f\tanh(f/2)}8},\frac12\right\}.
 \label{em:eq:switch-transient}
\end{align}
This expression is independent of the input sign and increases with
$\lambda\le\lambda(b)$. Taking the supremum therefore gives \cref{eq:switch-gamma-transient}.

\begin{figure}[b]
 \centering
 \includegraphics[width=\columnwidth]{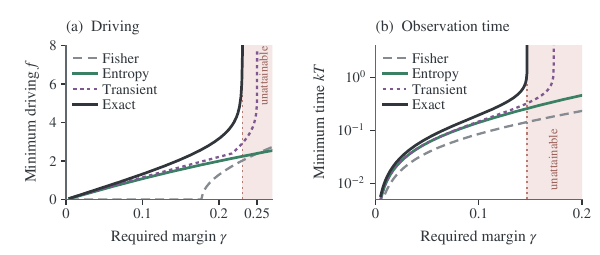}
 \caption{Minimum driving (a) and observation time (b) needed for copying and
 inversion at margin $\gamma$, with $b=0.5$, uniform preparation, and
 $x=\pm1$. We fix $kT=0.25$ in (a) and $f=1.5$ in (b).
 Black gives the exact minimum; other curves give lower bounds.
 Shaded margins cannot be reached.}
 \label{fig:switch-margin-thresholds}
\end{figure}

\Cref{fig:switch-margin-thresholds} shows that achieving a larger switching margin requires stronger driving and more time.

\end{document}